\documentclass[]{spie}  

\usepackage{amsmath,amsfonts,amssymb}
\usepackage{booktabs}
\usepackage{graphicx}
\usepackage[colorlinks=true, allcolors=blue]{hyperref}
\usepackage[T1]{fontenc}
\usepackage{enumitem}

\usepackage{wrapfig}

\usepackage{array}
\usepackage{makecell}

\title{CCAT: The 410 GHz camera module for FYST -- design and testing of the MKID focal plane.}

\author[a,b,c]{Scott C.\ Chapman}

\author[d]{Jordan Wheeler}
\author[b]{Anthony Huber}
\author[d]{Jason Austermann}
\author[d]{James A.\ Beall}
\author[b,c]{James Burgoyne}
\author[e]{Max Chapman}
\author[b]{Jesslyn Devina}
\author[f]{Douglas Henke}
\author[d]{Johannes Hubmayr}
\author[e]{Dylan Lockyer}
\author[g]{Michael Niemack}
\author[h]{Tilak M.\ Patel}
\author[a]{Colin Ross}
\author[b]{Douglas Scott}
\author[i]{Adrian Sinclair}
\author[h]{Eve Vavagiakis}
\author[d]{Michael R.\ Vissers}
\author[b]{Ruixuan (Matt) Xie}

\affil[a]{Dept. of Physics and Atmospheric Science, Dalhousie University, Halifax, BC, Canada}
\affil[b]{Dept.\ of Physics and Astronomy, University of British Columbia, Vancouver, BC, Canada}
\affil[c]{Dept. of Physics and Astronomy, University of Victoria, Victoria, BC, Canada}
\affil[d]{Quantum Sensors Group, National Institute of Standards and Technology, Boulder, CO, United States}
\affil[e]{Department of Engineering, McGill University, Montreal, QC, Canada}
\affil[f]{NRC Herzberg Astronomy \& Astrophysics Research Centre, Victoria, BC, Canada}
\affil[g]{Department of Physics, Cornell University, Ithaca, NY 14853, USA}
\affil[h]{Department of Physics, Duke University, Durham, NC 27710, USA}
\affil[i]{William H.\ Miller III Department of Physics and Astronomy, Johns Hopkins University, Baltimore, MD, USA}

\authorinfo{Further author information: Send correspondence to S.C.C.: E-mail: scott.chapman@dal.ca}

\begin{document} 
\maketitle

\begin{abstract}
Prime-Cam, the primary first-light instrument for the Fred Young Submillimeter Telescope (FYST) developed by the Cerro Chajnantor Atacama Telescope (CCAT) Collaboration, will accommodate seven modules. Here we describe the off-central 410\,GHz imager/polarimeter. The 410 GHz instrument is a camera module for CCAT funded by the Canadian Foundation for Innovation, being developed as a collaboration between Dalhousie University, University of British Columbia (UBC), National Research Council (NRC) - Herzberg, and Duke University. With atmospheric loading in the 410\,GHz window being significantly higher than at 350\,GHz (but substantially lower than 850\,GHz), we assess four MKID test devices with varying inductor volume for performance at 410\,GHz. We propose a design for an array of $\sim$6,700 horn-coupled TiN MKIDs optimized for use at 410\,GHz, with a planned $\sim$20,000 MKIDs over three arrays, exploring mapping speed versus detector number. We test the four MKID devices optically and assess optimal $Q_{\rm i}$/$Q_{\rm c}$ and responsivity for the 410\,GHz atmospheric window atop Cerro Chajnantor. The detectors will be designed in frequency to be efficiently readout with a second generation, two octave readout (based on the Xilinx RFSoC board), with over 4000 detectors per board.
\end{abstract}

    
\section{Introduction}
The  Fred Young Submillimeter Telescope (FYST)  at the Cerro Chajnantor Atacama Telescope (CCAT) observatory is a 6-m aperture submillimeter (sub-mm) to millimeter (mm) wave telescope, situated near the Cerro Chajnantor summit high in the Atacama Desert in Chile, and completed in April 2026. FYST is sited 5600 meters above sea level, providing a high, dry site suitable to negate much of the atmospheric effects, such as absorption by atmospheric water vapor, which plague terrestrial infrared observations \cite{P1}.
FYST’s extremely wide field of view (FoV) crossed-Dragone design will enable far faster mapping than existing facilities (Figure 1: \cite{Niemack:16}) The off-axis 6-m telescope design\cite{Niemack:16} achieves very low, $<$1\%, emissivity to take advantage of the Cerro Chajnantor site for sub-mm/mm wave astrophysics. The high surface accuracy, half-wavefront error of $\sim$10.7$\mu$m, ensures excellent sub-mm sensitivity, crucial for observations at the highest frequencies.

Prime-Cam, the primary first generation instrument on FYST, incorporates seven independent instrument modules to accommodate a range a scientific goals (Fig.$~$\ref{fig:FYST/Cam}: Right).
Similar cameras have been previously developed for astronomical observations \cite{P2,Eve}, but the unique capability of exploring the 410\,GHz band makes this instrument crucial, as there are no immediate proposals for another instrument with this capability\cite{Scott}.
The off-central 410\,GHz imager/polarimeter is a module in Prime-Cam funded by the Canadian Foundation for Innovation, being developed as a collaboration between Dalhousie University, UBC, NRC-Herzberg and Duke University. 
FYST will provide unrivaled mapping speeds at this frequency (Fig.~1) \cite{Stacey}, enabling  observations beyond the confusion limited depth of previous or planned space-borne observatories \cite{primecollaboration2021ccatprime}. 

\begin{figure}
\centering
\includegraphics[width=0.99\linewidth]{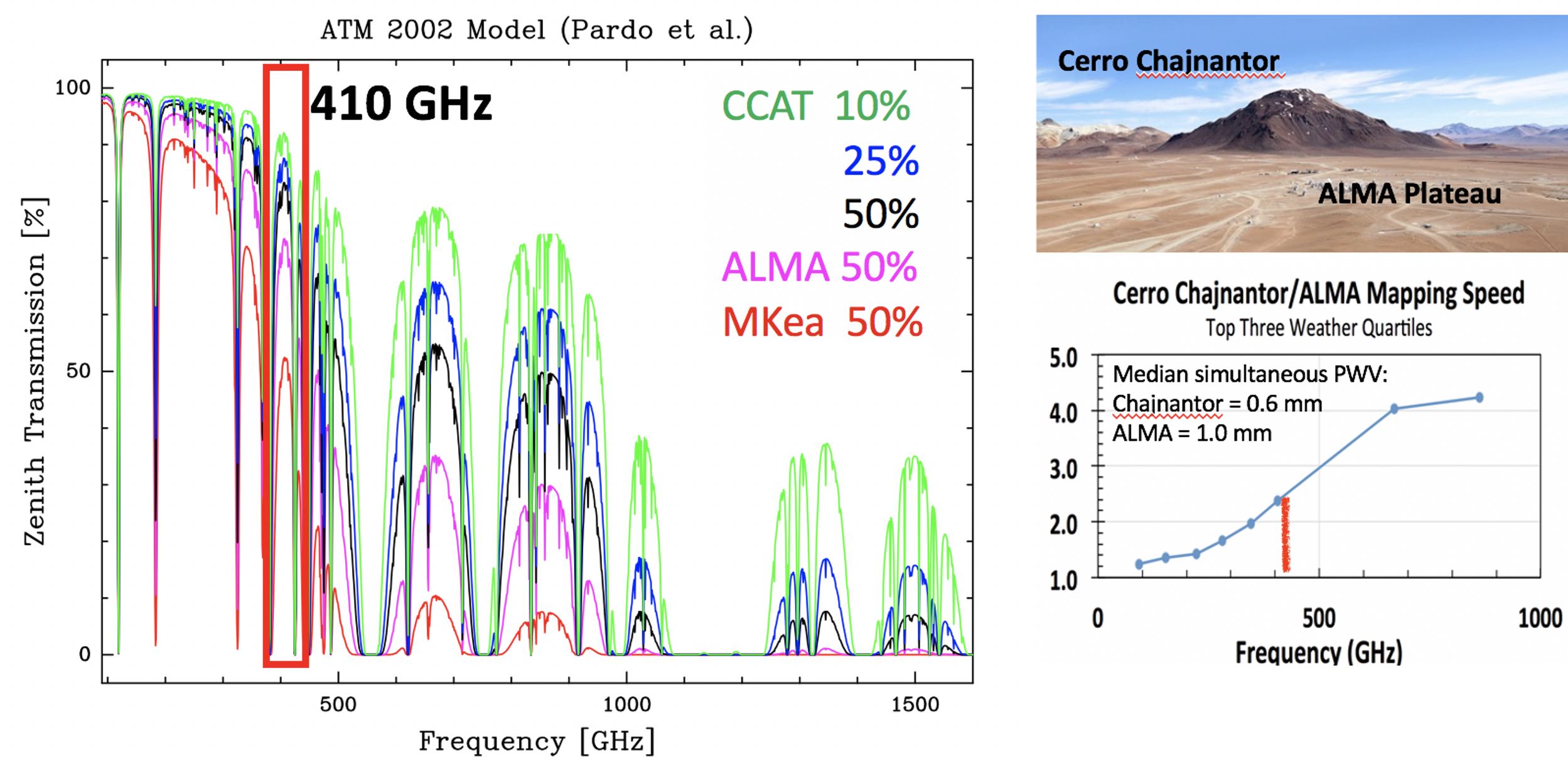}
\caption{Left: Atmospheric transmission at the CCAT site, Cerro Chajnantor, identifying the exquisite 410\,GHz transparency compared to the ALMA and Mauna Kea sites.  Right Top: a photo of the site, and Right Bottom: the improvement in sub-mm mapping speed expected at 5600\,m relative to that of the ALMA Plateau, highlighting 410\,GHz}
\label{fig:FYST/Cam}
\end{figure}

Of the instrument modules currently being developed for the Prime-Cam receiver, this second highest frequency 410\,GHz module presents  challenges in optical design, coupling, detection, and readout.
About 20,000 polarization-sensitive, lumped-element microwave kinetic inductance detectors (MKIDs) will be incorporated into the 410\,GHz module.
This work presents key aspects of the detector design and solutions to the challenges of efficient optical coupling and a multi-octave readout band with 1000 resonators per channel to be read out using an Xilinx ZCU111 RFSoC board \cite{Adrian}.

The wavelength coverage, sensitivity, spatial resolution, and large FoV of Prime-Cam on FYST allow for a set of wide-area surveys, between 5 and 15,000\,deg$^2$, to be conducted in order to address many scientific goals\cite{primecollaboration2021ccatprime}. The 410\,GHz instrument will play a key role in many of the science cases, complementing the higher frequency 850\,GHz module also under development(Chapman et al.\ 2022\cite{Scott2022}, Burgoyne et al.\ 2026, this volume, Huber et al.\ 2026, this volume):\\ 
1. Directly trace the evolution of dusty-obscured star formation in galaxies since the epoch of galaxy assembly\cite{chapman2005}, starting $>$10 billion years ago. Comparable to ALMA's band-8, the Prime-Cam 410\,GHz module offers a novel frequency band for tracing dusty galaxy evolution, never before widely probed in surveys. The 410\,GHz band will offer a powerful constraint on galaxy dust temperatures and far-infrared luminosities \cite{kovacs06,chapman2003}\\
2. Constrain dark energy and feedback mechanisms by measuring the physical properties and distribution of galaxy clusters via the Sunyaev-Zeldovich (SZ) effect on the CMB.\\
3. Enable more precise constraints on inflationary gravitational waves and light relics by measuring polarized CMB foregrounds and Rayleigh scattering. The addition of the 410\,GHz channel has been shown to significantly reduce the bias in these constraints over 350\,GHz and 850\,GHz alone \cite{primecollaboration2021ccatprime}\\
4. Monitor and search for time-dependent events: a wide variety of transient sources are becoming a large focus in submm and mm surveys.\\
5. Determine the role of magnetic fields in star formation by measuring Galactic polarization at an intermediate frequency between the typical baselines of 350\,GHz and 850\,GHz.

\begin{figure}
\centering
\includegraphics[width=0.52\linewidth,height=0.475\linewidth]{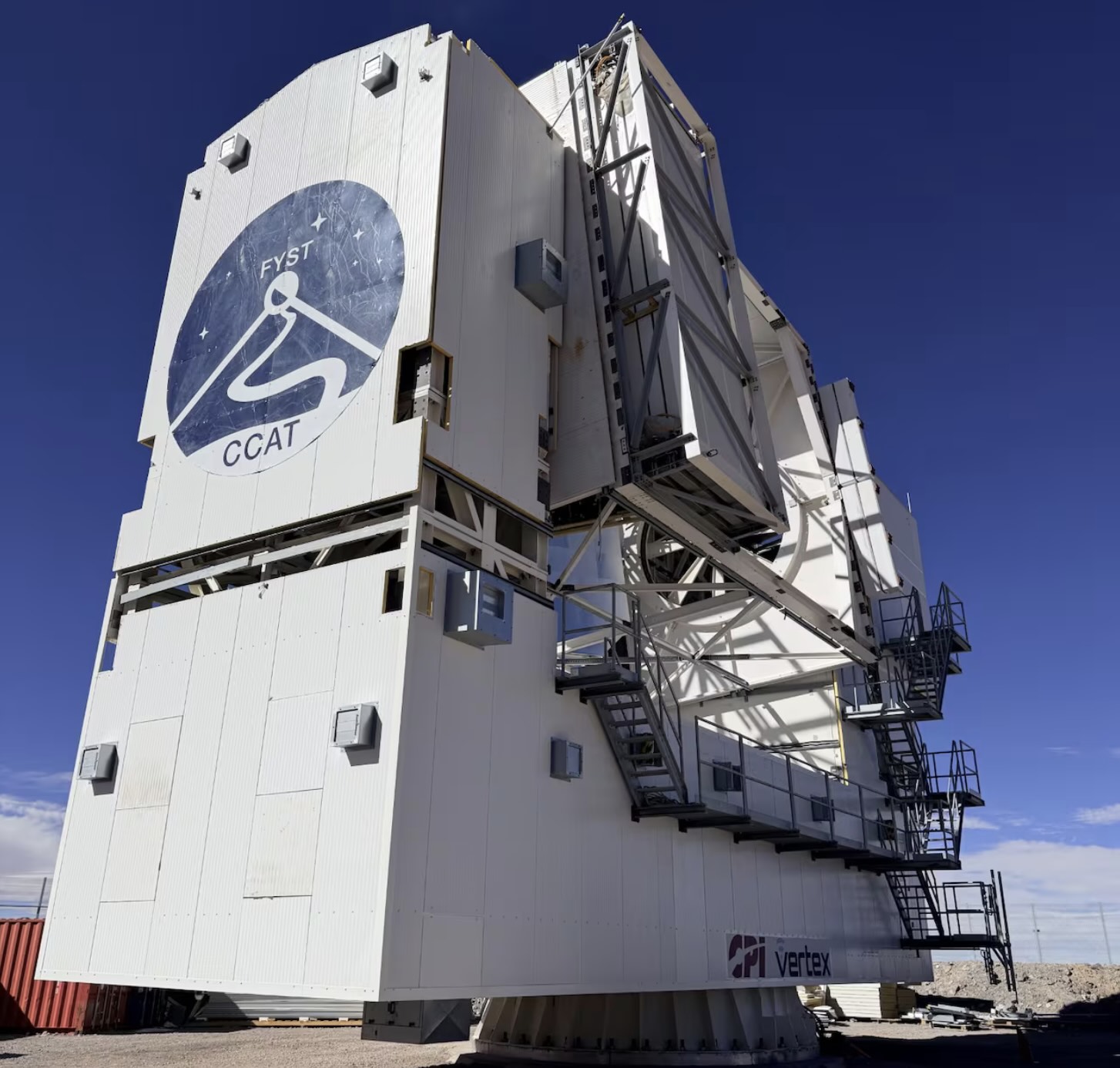}
\includegraphics[width=0.47\linewidth,height=0.475\linewidth]{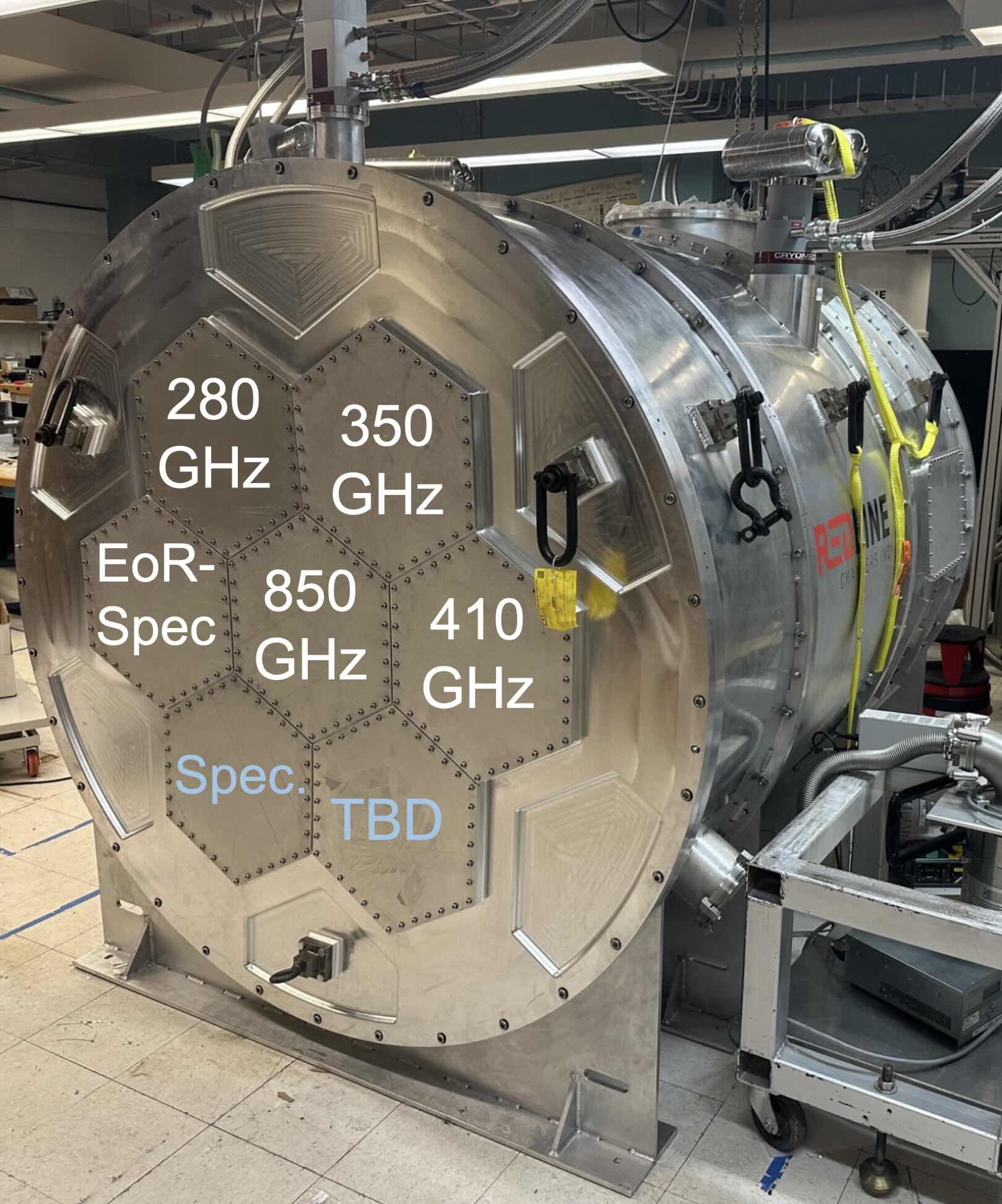}
\includegraphics[width=0.66\linewidth]{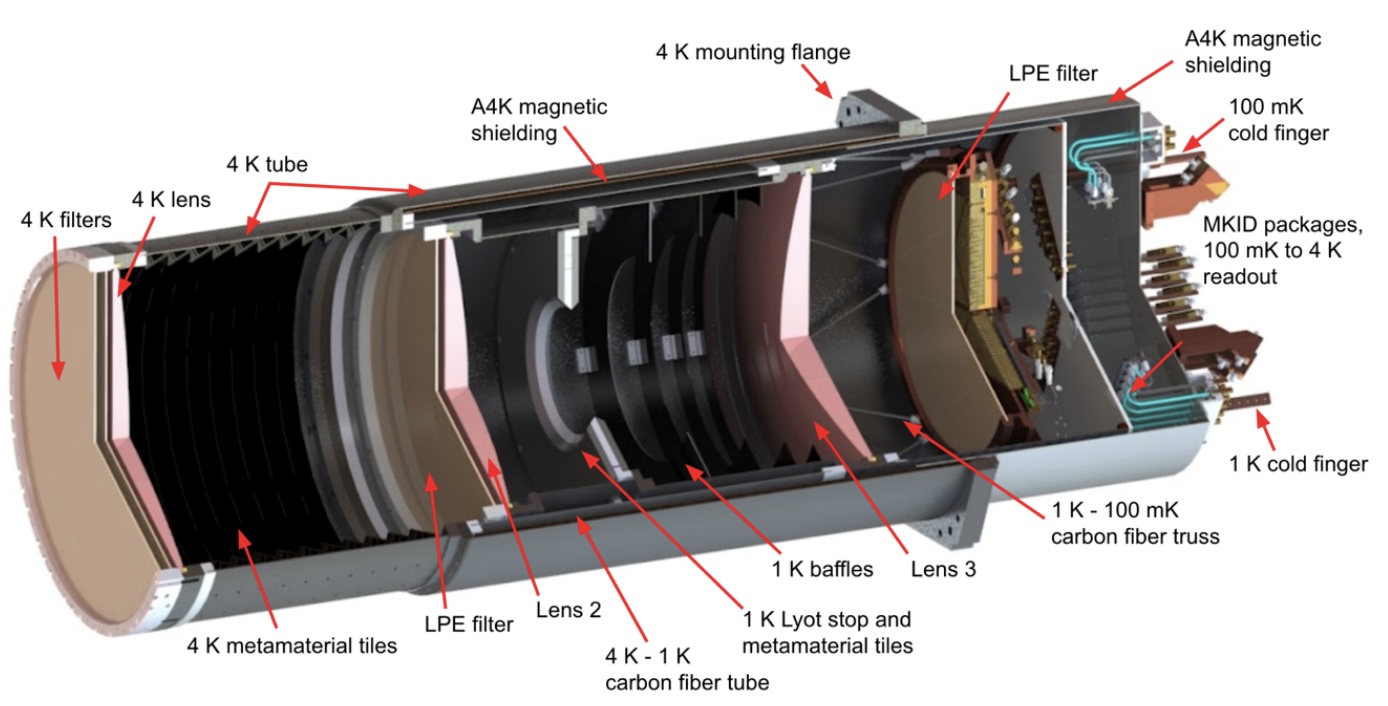}
\caption{{\it Top Left:} FYST sited at an elevation of 5600 m (photo credit: Margaret Chapman). {\it Top Right:} Configuration of instrument modules within Prime-Cam \cite{vavgiakis_mmuniverse}. 
{\it Bottom:} Schematic of the planned 410\,GHz optics tube, with location of the MKID arrays described herein indicated at right. The 410\,GHz module is a close replication of the 350\,GHz module\cite{Duell:24}.
}
\label{fig:FYST/Cam}
\end{figure}

\section{Mapping speed and 410\,GHz module design considerations}
\label{sec:mapping}

At 410 GHz, the 1.3$^{\circ}$ FoV of the FYST camera module coupled with the atmospheric transmission advantage at 5600\,m (Fig.~1), implies more than a factor 300 improvement in mapping speed achievable compared to a similar instrument on the JCMT, given the lower and wetter site on Mauna Kea and the smaller 8$'$ field of view.

The approximately 0.4-m diameter optics tube will illuminate three 150-mm detector wafers. The optics tube size provides an unobstructed 1.3$^\circ$ diameter FoV and keeps the size of the entrance window manageable.
Patel et al.\ 2026 (this volume) present optical design considerations for the 410\,GHz camera module. To enable rapid deployment and manage costs, the 410\,GHz module replicates the three-lens 350\,GHz module optical design (Keller et al.\ 2026, this volume) -- see Fig.~2. Patel et al.\ 2026 show that such a three-lens optical design only slightly diminishes the optical performance at 410\,GHz across the field, enabling near diffraction limited Strehl ratios over most of the field covered by the three detector arrays.

\begin{figure}
\centering
\includegraphics[width=0.99\linewidth]{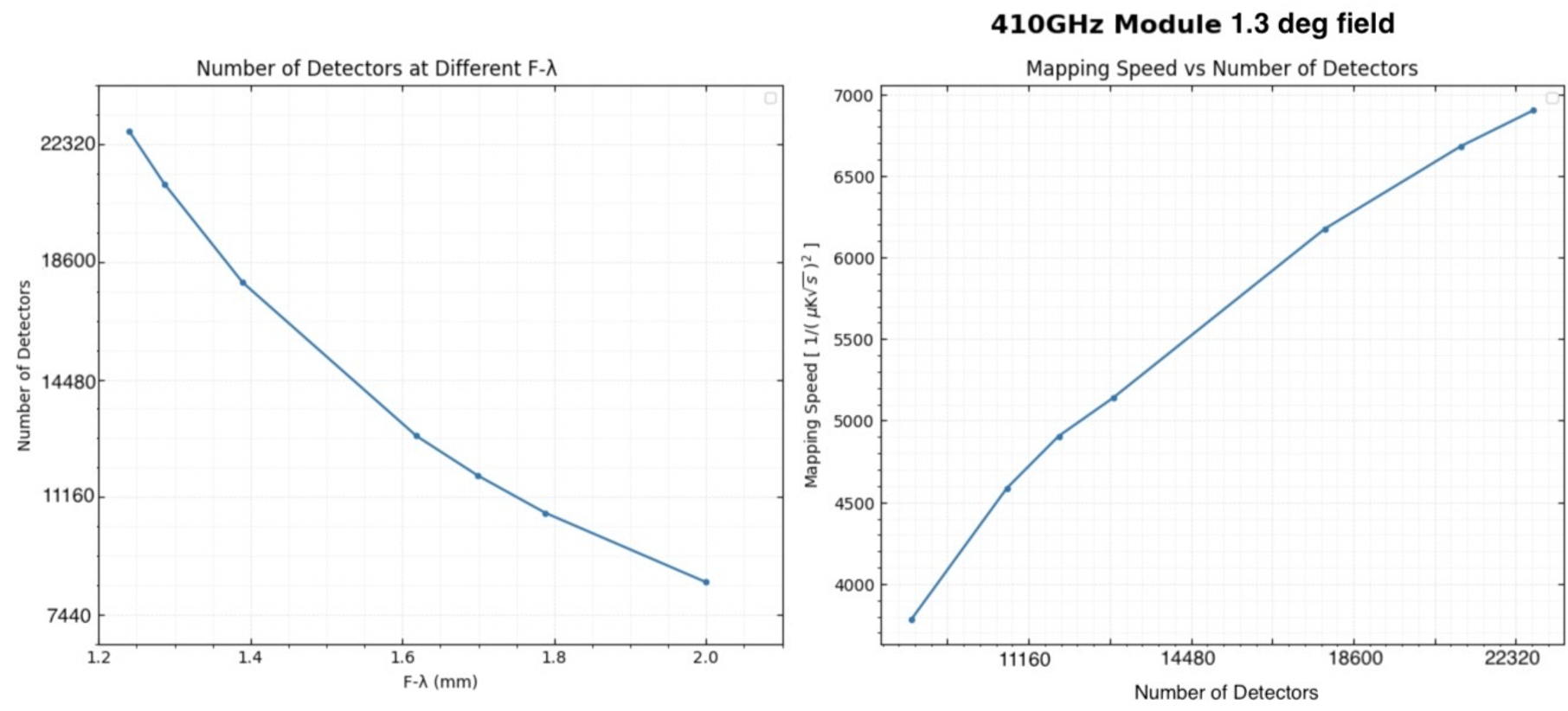}
\caption{Left: Number of detectors required to fill the candidate  camera designs with  fields-of-view  1.3$^{\circ}$  as a function of F-$\lambda$ pixel spacing. The range of required detectors spans that which is feasible using single layer lithographic fabrication of the capacitors, inductors and feed lines (see section~3). Right: Mapping speed as function of number of detectors  for point sources (similar for extended sources). 
For detector numbers ranging from 10,000 to 20,000, there is a marked $\sim$60\% improvement in mapping speed due to the  advantageous gain on the spillover/aperture efficiency curves for horn-coupled detectors.}
\label{fig:FYST/Cam}
\end{figure}

Here we present a study of the pixel density and mapping speed, providing a framework to consider the MKID and readout design tradeoffs for 410\,GHz.
Fig.~3 shows the number of detectors required to fill the 1.3$^{\circ}$ diameter FoV as a function of the F-$\lambda$ pixel spacing, with each pixel requiring two detectors in order to sample both polarizations. The key trade-offs are shown in Fig.~2 where the mapping speed is assessed as a function of the number of detectors filling the field.
For horn-coupled detectors, the formula below shows how the normalized sensitivity is affected predominantly by the spillover effciency (for extended sources) or aperture effciency (for point sources). The effects on the mapping speed are similar from both the spillover and aperture effciencies \cite{Griffin}.

$${\rm Sens.} \propto \frac{1}{\rm \sqrt{{ Throughput \times No.\ dets \times Effic.(spill\ or\ aper) }}}\ \ ,$$
where mapping speed is inversely proportional to the square of the sensitivity.

The detector coupling efficiency to the point spread function is important when measuring the flux density of point sources. 
The realistic maximum feedhorn aperture efficiency is 0.7 for an aperture of 2\,F-$\lambda$. However increasing the density of horns (pixels) to $<$2\,F-$\lambda$ increases the overall mapping speed for a fixed FoV.
The aperture efficiency at each F-$\lambda$ is taken from \cite{Griffin}.
When observing extended sources, the signal power absorbed by a detector is produced by the astronomical sky brightness.
The spillover efficiency of the horns is the dominant factor affecting mapping speed for different F-$\lambda$ spacings, and is dependent on the horn edge taper. 
The edge taper values for F-$\lambda$ between 1 and 2 are taken from \cite{Griffin}.
Compared to a baseline of 10,000 MKIDs adopted in the lower frequency modules and a 1.3$^{\circ}$ FoV \cite{Choi:22}, Fig.~3 shows more than 60\% improvement in mapping speed if   a design can be accommodated with $\sim$20,000 MKIDs, providing an F-$\lambda$$\sim1.3$. As with the optical design considerations above, the most straightforward path to a fast track implementation of the 410\,GHz module involves replicating the RF-cryogenic architecture of the 350\,GHz module (Keller et al.\ 2026), which has 18 feed lines. With the current Gen-2 two-octave readout (\cite{Xie:24}, Xie et al.\ 2026 this volume), this would limit the detector count to about 18,000 MKIDs.

\section{MKID Design}
\label{sec:design}

Designing the MKIDs for the lower readout frequencies of the RFSoC (300--1,200\,MHz for this module)  presents significant challenges in achieving the high quality factor ($Q$) and signal-to-noise ratio (SNR) necessary for scientific observations, while simultaneously fitting all the resonators into the available bandwidth. To meet these demands, the module employs a two-octave design, which carefully balances resonator spacing and readout efficiency, ensuring optimal performance across a wide frequency range without sacrificing sensitivity.

The key parameters explored here for the 410\,GHz MKIDs include 
testing different volumes of the inductor both in terms of line width and deposition thickness to tune the sensitivity.
The designs include a feature to reduce the inductance across a portion of the detectors by shorting pairs of inductor lines to allow MKIDs to be tuned across four distinct bands across the readout range, all with minimal impact to the responsivity of the detector. This shorted inductor technique has been implemented in the 850\,GHz design \cite{Huber:24}.
The resonators are coarsely tuned via the inductance shorts, and finely tuned by etching away small portions of the interdigital capacitors (IDCs)\cite{Huber:24}.
What follows represents the latest developments towards the design and fabrication of a  densely-packed far-infrared lumped element MKID array for the 410\,GHz polarimeter.


The detectors chosen for the 410\,GHz module of Prime-Cam are lumped-element kinetic inductance detectors.
While aluminum-based meander inductors have been employed in other modules due to their performance benefits, particularly in mitigating $1/f$ noise (see Ref. \cite{Duell:24} for details), they present challenges when dealing with the higher loading requirements of the 410\,GHz band. Specifically, the increased inductor volume needed to maintain ideal optical coupling while handling this higher loading becomes difficult with aluminum.
To address this, titanium nitride (TiN) has been selected as the superconducting material 
for its higher kinetic inductance and resistivity, which are advantageous under the higher optical loading conditions encountered at 410\,GHz. These 
properties facilitate the optimization of efficient optical coupling while maintaining the required inductor volume for optimal detector performance. At the  higher atmospheric loading of 410\,GHz through 850\,GHz, the $1/f$ noise in TiN has been shown to be subdominant to photon noise \cite{Huber:24}.
Since the 410\,GHz module functions as a polarimeter, each pixel requires two MKIDs, with orthogonal polarization to accurately measure polarized light. This dual-polarization design, while increasing pixel complexity, is critical for achieving the module's scientific goals, particularly in studying the polarization of submillimeter astronomical sources with high sensitivity.

\subsection{Resonator Design}
\label{sec:design:Resonator}

The primary challenge of the 410\,GHz detector array is the relatively dense-packing of the detectors on the focal plane, similar to the higher frequency 850\,GHz array \cite{Huber:24}.
One of the key constraints is that the capacitors occupy a significant portion of the overall MKID footprint, making it crucial to optimize their design in order to minimize pixel size while still maintaining the required performance.
This balance between compactness and functionality is central to achieving the necessary pixel density.
The proposed pixel pitch for the 410\,GHz module is 2.0\,mm.
The resonance is inversely proportional to $\sqrt{L_{\rm tot}C}$, where $L_{\rm tot}$ is the total inductance and $C$ is the capacitance, we see that to reduce the size of the capacitor, either the interdigitated capacitor (IDC) fingers and gaps must decrease, which would potentially increase TLS noise \cite{Noroozian:09}, or the total inductance must increase in order to keep the same range of resonant frequencies, which is limited by the readout band.

\begin{figure}%
    \centering
   \includegraphics[width=0.45\linewidth]{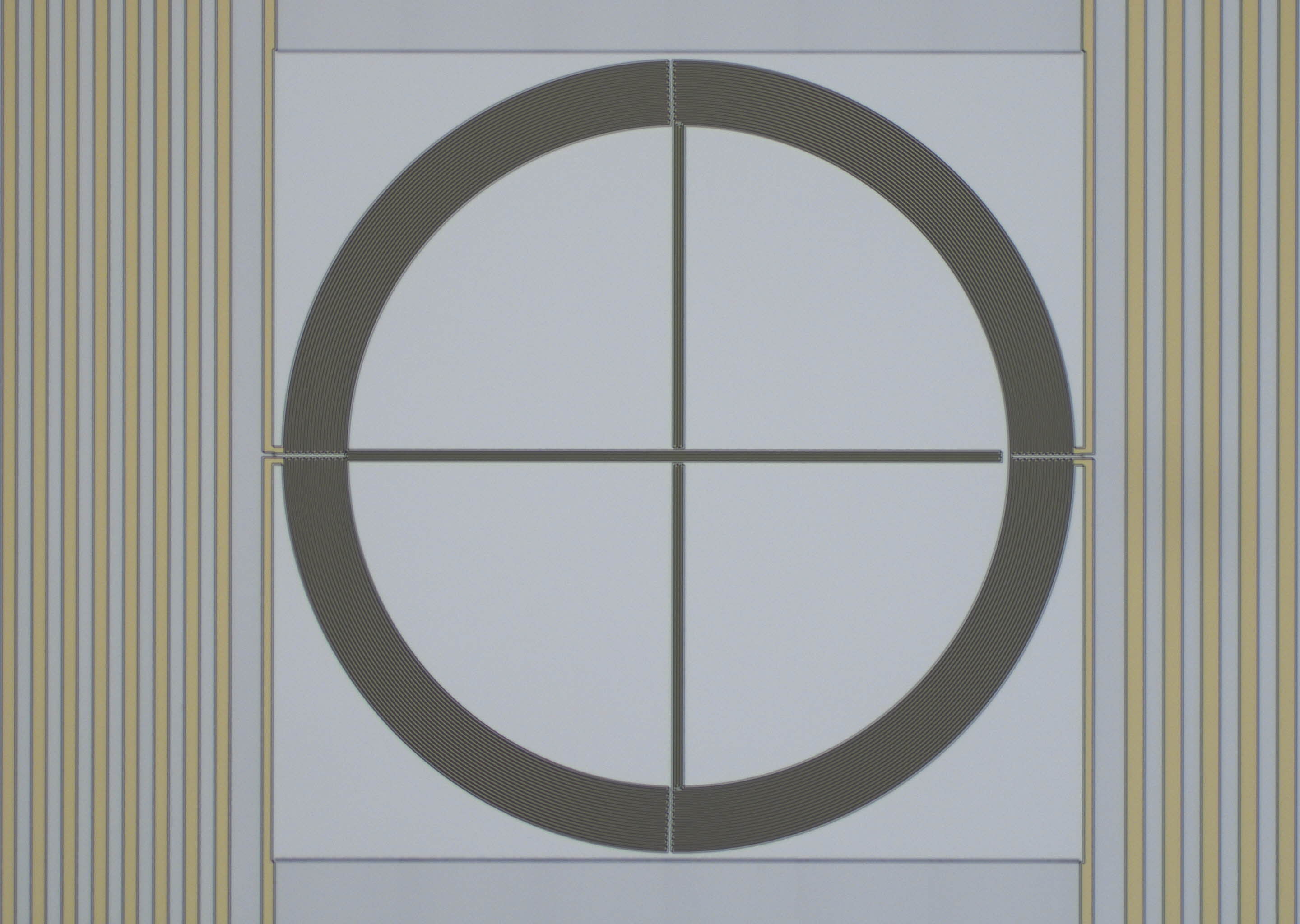}%
    \hspace{0.3cm}%
    \includegraphics[width=0.065\linewidth,trim={0 4.5cm 0 0 },clip]{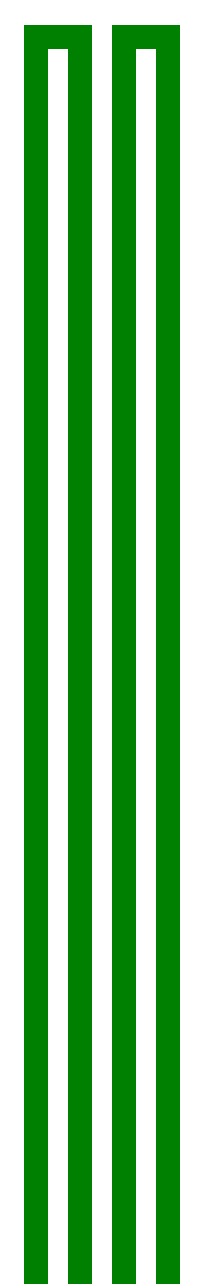}
    \includegraphics[width=0.065\linewidth,]{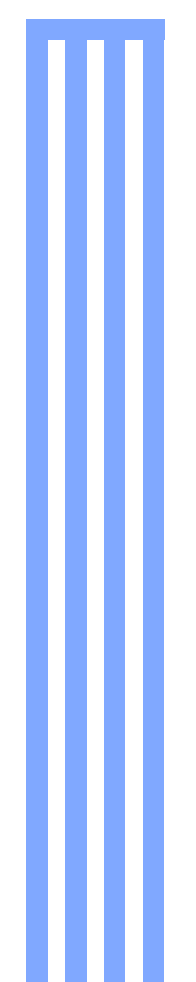}
     \hspace{0.3cm}%
\includegraphics[width=0.35\linewidth,height=0.325\linewidth]{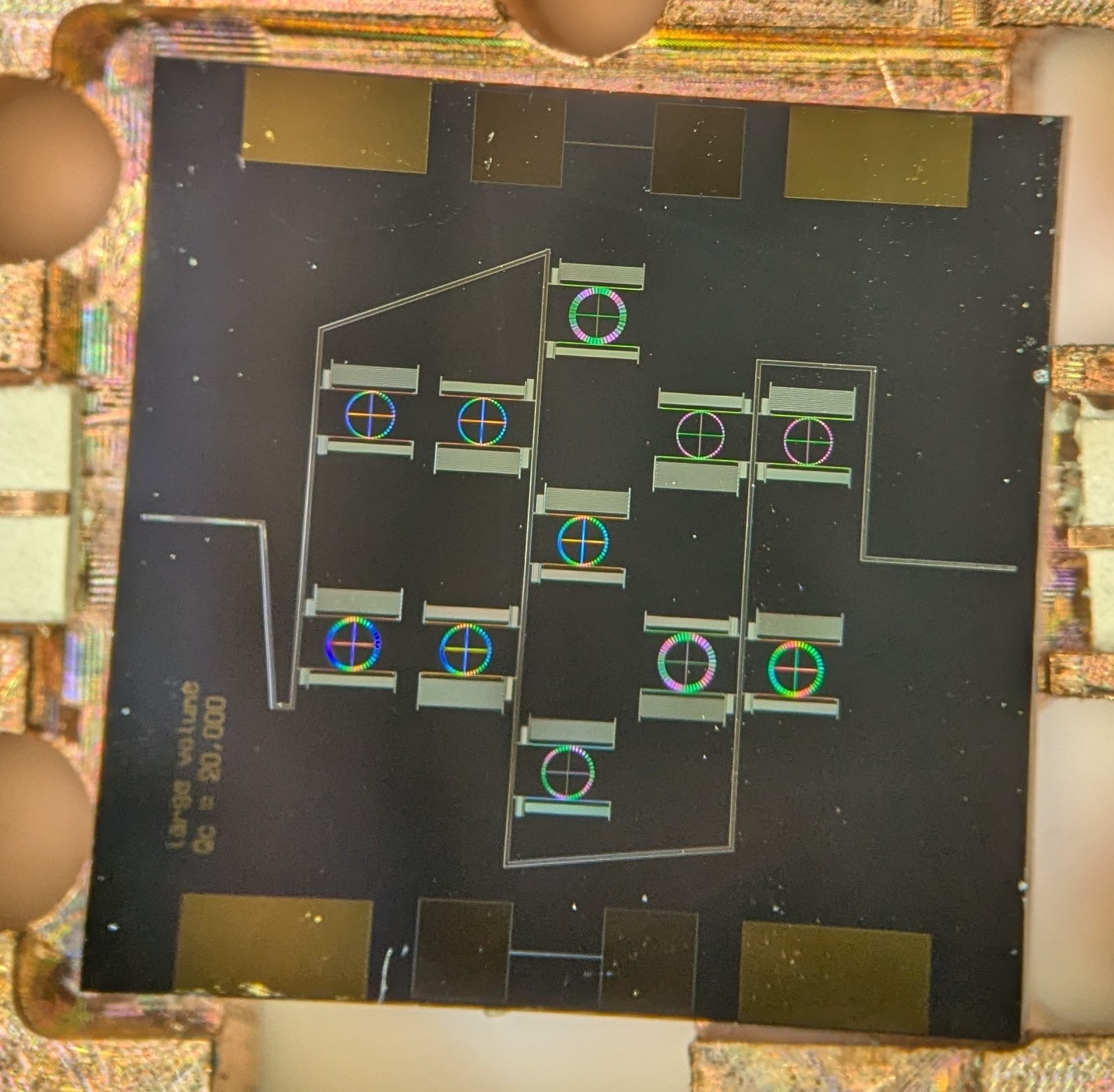}
    \includegraphics[width=0.51\linewidth]{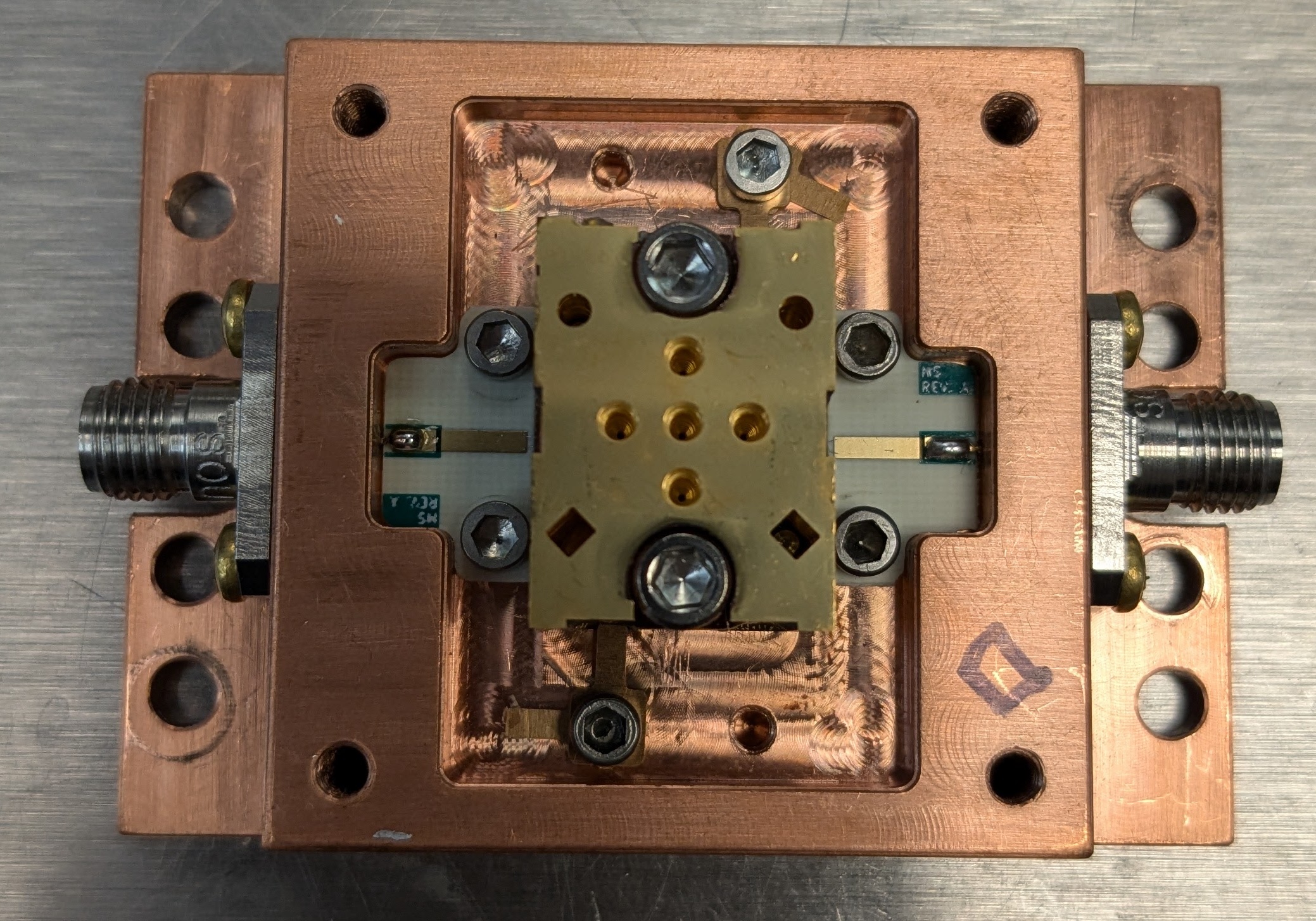}
    \includegraphics[width=0.47\linewidth]{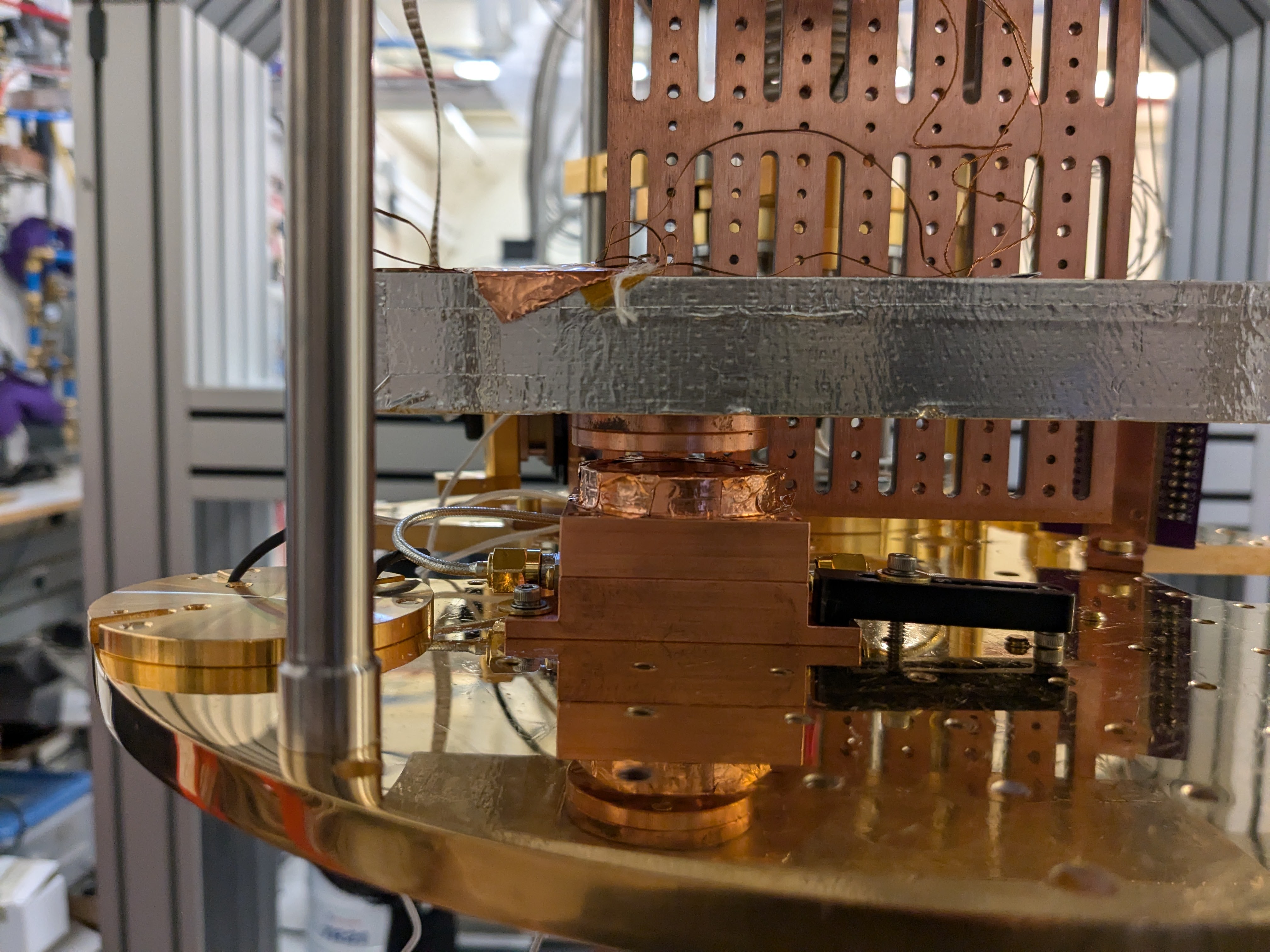}
    \caption{{\it Top Left:} Sample fabricated inductor  with two polarization-sensitive kinetic inductance detectors in a single pixel. Each inductor consists of dark inductance around the arc (defining the circular shape) and optically active inductor (seen as the ``crosshair''). 
    The directions of the optically active inductor lines determine the polarization sensitivity of the detectors The diameter of the dark inductance ring is of the order of 1 mm. {\it Top Middle:} Zoomed image of sample designs of non-shorted (green) and shorted (blue) inductor lines from the optically active region. The inductor lines are 5\,\textmu m wide with 5\,\textmu m gaps between lines. {\it Top Right:} A test device analyzed in this work with different volume inductors in each MKID. {\it Bottom Left:} The test device mounted behind machined test feedhorn block.  {\it Bottom Right:} The test setup in the NIST DR.
    }
    \label{fig:InductorExample}%
\end{figure}

From Fig.$~$\ref{fig:InductorExample}, as described in \cite{Huber:24}, using the different thickness and line widths, the ratio of active inductance (absorber) to dark (or unilluminated) inductance can be modified and tuned to optimize the resonator design.
Here dark inductance is defined as inductance which contributes to the resonant frequency but not to the absorption of photons.

The 
optical responsivity of TiN
  MKIDs is defined as \cite{Gao, Mauskopf:18}:
\begin{equation}
	\frac{\text{d}f_0}{\text{d}P_{\rm abs}} \simeq \frac{\alpha f_0}{2N_0\Delta_0} \bigg(1 + \sqrt{\frac{2\Delta_0}{\pi k_{\rm B}T}}\bigg) \frac{\eta\tau_{\rm eff}}{\Sigma}
	\label{dfDP}
\end{equation}


\noindent respectively, where $\alpha = L_K/L_{\rm tot}$ is the kinetic inductance fraction, $L_{\rm K}$ is the kinetic inductance, $\Delta_0$ is the binding energy for a single electron, 
$N_0$ is the material-dependent single spin density of states at the Fermi energy, $\eta$ is the internal quasiparticle generation efficiency, $\tau_{\rm eff}$ is the energy relaxation time, and $\Sigma$ is the superconducting volume in the detector.
Thus the optimization of a KID depends on  several parameters, many of which are material dependent.

The 410\,GHz module will experience relatively high loading during operation ($\sim$10-20 pW).
From Eq.~\ref{dfDP}, the effect of loading on the responsivity can be suppressed by increasing the superconducting volume.
This takes the form of increasing line widths of inductor lines or increasing the film thickness, both of which directly impact the kinetic inductance as it is proportional to both the width and thickness.
Balance must be struck between the appropriate value of inductance, responsivity, and film thickness.
A typical TiN film used for MKID fabrication consists of a TiN layer sandwiched between two layers of Ti, known as a trilayer film \cite{Vissers:13}; this work uses a septalayer design (TiN/Ti/TiN/Ti/TiN/Ti/TiN) {to better tune the film parameters, providing} another variable for optimizing for footprint size and resonator volume while maintaining the desired responsivity and impedance matching.  This allows us to probe a large range in volumes to find the most suitable design for 410\,GHz operation.

\subsection{Shorting Inductor Lines}
\label{sec:design:Inductor}

The typical inductor (Fig.$~$\ref{fig:InductorExample}) design from Huber et al.\ \cite{Huber:24} consists of dark inductance around the arcs (circular shape) and optically active inductance under the feedhorn (cross-hair shape).
To reduce the inductance in a detector a technique was developed based on shorting pairs of inductor lines \cite{Huber:24}, which allows for a two-octave design without complications (e.g., parasitic capacitance constraining the overall capacitance) resulting from varying capacitors by a factor of 4. This also helps avoid the spatial constraints on the pixel size being dominated by the resonator capacitors.

Shorting inductor lines means coupling adjacent lines into bundles, providing the same optical coupling and decreased kinetic inductance since the effective number of squares has decreased.
When $N$ lines are shorted together, the result is a decrease in the kinetic inductance proportional to $1/N^2$.
The geometric inductance, however, is at most minimally impacted by shorting detectors; it is critical that $\alpha$ be maximized when shorting inductors, else the impact of shorting inductor lines will be diminished as the geometric inductance contributes more to the total inductance.
%
There is  no significant change to the 
submillimeter impedance or sheet resistance, preserving responsivity and optical coupling efficiency.

For this work,  two lines are shorted together at a time, meaning resonators will have kinetic inductance values of $L_{\rm K}$ and $L_{\rm K}/4$ for non-shorted and shorted inductors, respectively.
This allows for a coarse tuning of the resonators across the readout bandwidth by having the inductor lines shorted or unshorted, and then  the resonators can be finely tuned via the interdigitated capacitors.
This process allows for the resonators to be much smaller than they would need to be for lower frequency designs, enabling the tight pixel pitch required for the 410\,GHz module {(2.0\,mm)}.
In our fabricated designs tested here, the lowest frequencies resonators have an inductance $\sim$200\,nH which is about 5 to 7 times the inductance of other Prime-Cam module detectors \cite{Duell:24,Wheeler:22}.
This in turn allows for the capacitors to be 5 to 7 times smaller.  

\begin{table}
    \centering
    \caption{Simulated resonant frequency range (MHz) of the 410\,GHz array based on the two IDC finger sizes and either shorted or non-shorted inductors.}
    \begin{tabular}{|r|cc|}
        \hline & 5 $\text{$\mu$}$m IDC & 12 $\text{$\mu$}$m IDC \\ \hline
        Non-Shorted & 300 - 435 & 435 - 635 \\
        Shorted & 635 - 925 & 925 - 1,350\\ \hline
    \end{tabular}
    \label{tab:FrequencyPlan}
\end{table}



The final MKID array designs at 410\,GHz will be based around two different inductor designs, which are identical except for two lines being shorted together.
Target values for the inductance are 200\,nH and 50\,nH, which assumes that the kinetic inductance dominates all other sources of inductance.
Note that in some simulations a geometric inductance as high as $\sim$10\,nH was found for models of potential KIDs.
Disregarding the geometric inductance, the target inductances were set as constants and the available ranges in capacitor size were then explored under the assumption that the overall footprint of the pixels was minimized; the simulations were essential in determining the ideal size of the capacitor fingers.
An IDC ideally has enough fingers to be able to tune the resonator across its given band, wide enough fingers to reduce TLS noise as as much as possible, and must maintain the requisite small footprint. 
Given that a typical lumped element KID operates around 1\,GHz, the 300\,MHz minimum frequency for the readout means the resonator designs require a large amount of inductance.

We  determine two different sets of capacitors which  allow for tuning across the entire planned 1-GHz readout band. The result is two sets of IDCs of differing widths, 5\,\textmu m and 12\,\textmu m, with identical spacing between lines.
The capacitance is then proportional to the number of lines of a given length.
Using the two values for the inductance and two sets of capacitors, the two octaves worth of resonators are easily obtained (See Table~\ref{tab:FrequencyPlan}).
The frequency spacing  assumes a spread of 0.15$\%$ between resonances, beginning at the readout minimum frequency of 300\,MHz.
The pixels are printed identically with minor variations, which helps achieve array uniformity and avoid design systematics. 
As described in \cite{Huber:24}, the line shorting provides a coarse tuning of the resonance, and the capacitors are relegated to fine tuning the resonators within the given band.

\begin{figure}[!t]
\centering
\includegraphics[width=.8\linewidth]{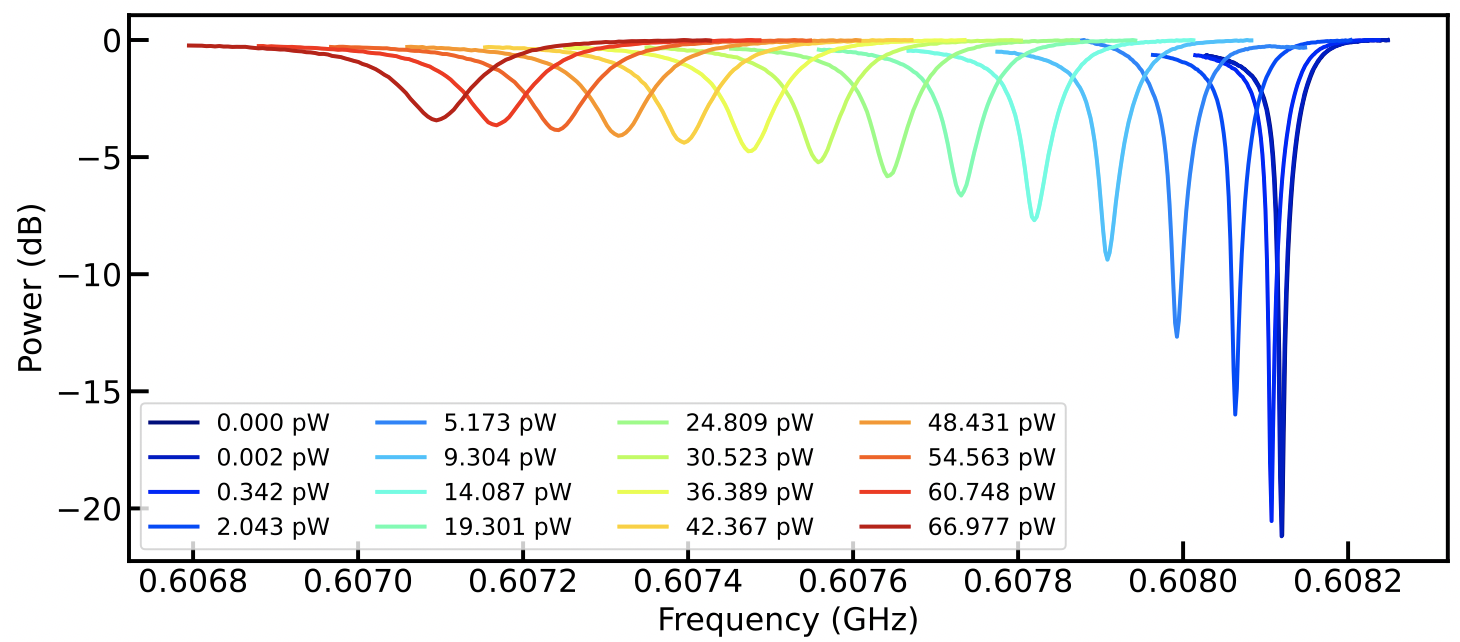}
\caption{Resonances with different loading for one of the 410\,GHz test devices with a dark resonance peak at 608.1\,MHz.}
\label{fig:TwoOctaveVNA}
\end{figure}

\begin{figure}%
    \centering
    \includegraphics[width=.96\linewidth]{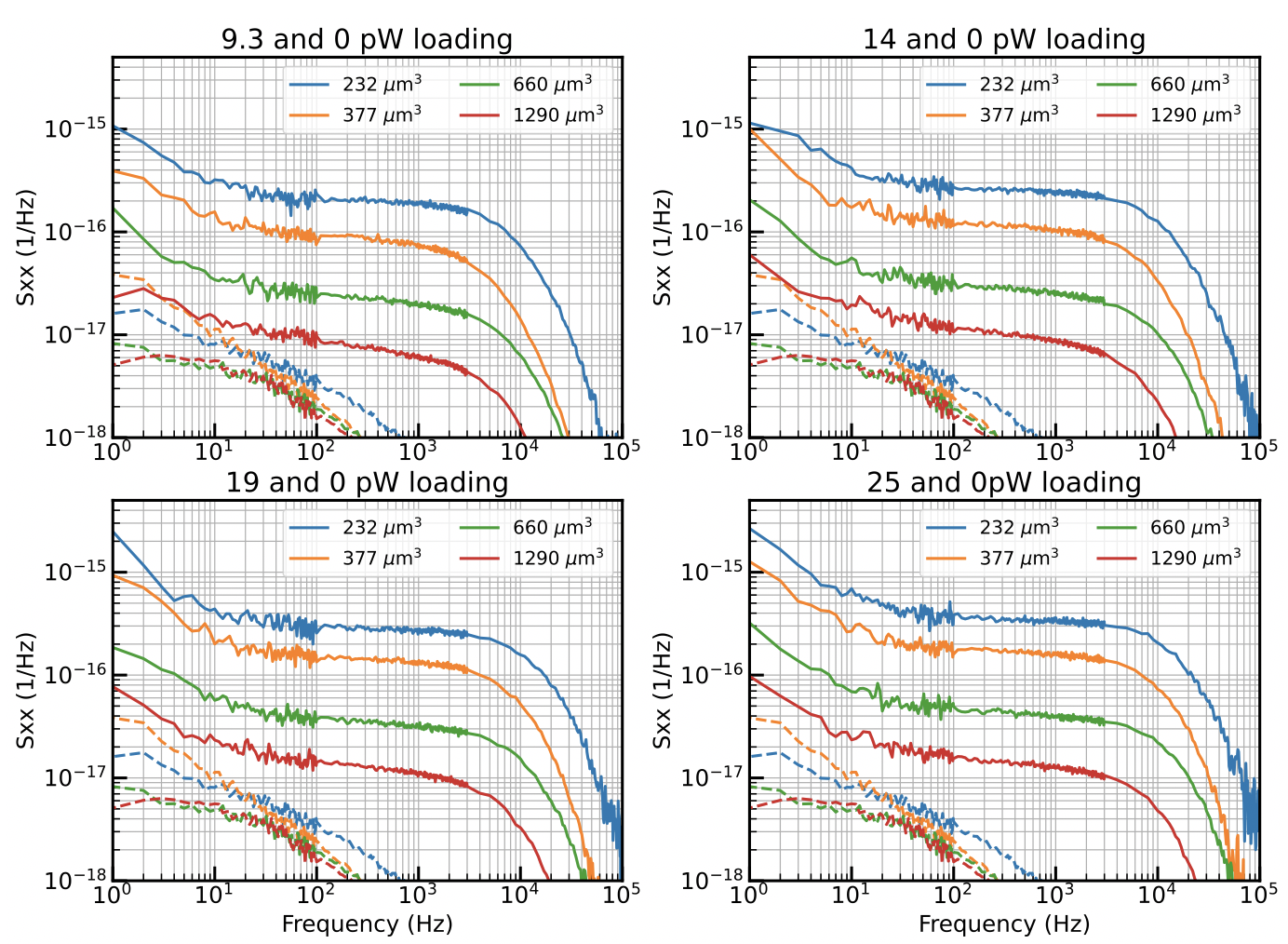}%
    \caption{Noise (in S$_{xx}$ units with increasing optical loading measured on the four different volume resonators (232 through 1,290\,$\text{$\mu$}$m$^3$). At lower loading, TLS noise is subdominant to photon noise, but still contributes significantly. At frequencies greater than roughly 2\,kHz the detector roll-off is observed.}%
    \label{fig:NEPThick}%
\end{figure}

\section{Experimental Setup}

Tests for the 410\,GHz prototypes were conducted in two cryostats, one at the National Institute of Standards and Technology (NIST - Boulder) - a Bluefors LD-400 
- and one at NRC-Herzberg Astronomy \& Astrophysics Research Centre (HAA) - a Bluefors LD-250.
Of these, only the NIST DR was used for optical testing due to the availability at the time of the requisite test kit including 
a calibrated cryogenic blackbody.
The 
NRC-HAA DR was used for dark tests, which included measuring the DC critical temperature and normal state resistivity 
of the septalayer film, performing bath temperature and readout power sweeps, and considering the characteristics of the two-octave design.
To achieve greater precision in the measurement of $Q$ values, additional magnetic shielding was included around the test chips.


Optical tests were performed  in the NIST DR.
A calibrated cryogenic blackbody was suspended from the still flange, with thermal offsets to prevent excess heating of the cryostat while achieving blackbody temperatures as high as 72\,K.
Two IR low-pass filters were attached to the output of the blackbody, and two more IR low-pass filters were used in the test box.
All filters were provided by Cardiff \cite{Ade}.
The low frequency edge of the band is defined by the diameter of the waveguide in the feedhorn block.
The test box was affixed directly below the cryogenic blackbody to ensure that the entire FoV of the detectors was filled by the blackbody.

\section{Results}

\subsection{Optical characterization of the 410\,GHz test detectors}

Figure$~$\ref{fig:TwoOctaveVNA} presents a VNA sweep of a test device under different optical loadings. 
Each of the 410\,GHz test resonators were fit to determine  parameters, such as $Q_{\rm i}$, $Q_{\rm c}$, and the time constant $\tau$.
The designs and application of the septalayer film resulted in dark $Q_{\rm i}$ ranging from  60,000 to 220,000 across the four test resonators.
The measured $T_{\rm c}$ for the TiN arrays was found to be roughly 850\,mK, which is ideal for the expected operational bath temperature of the 410\,GHz module focal plane (100\,mK).
The septalayer also yielded a measured normal resistance of the TiN, 31.5 $\Omega/\square$ and a kinetic inductance of 51.1 pH/$\square$.
The time constant for the TiN test detectors was found to be in the $\sim$0.2\,ms.

The Noise Equivalent Power (NEP) for the four volumes (232, 377, 660, and 1,290  $\mu$m$^3$) of resonators were measured, increasing the power on the black body from 0 to 67\,pW.
Results show that for all volumes there is a large shift in the noise level with small increases in loading levels, particularly at the lower limit of optical loading.
The measured noise properties (S$_{xx}$) of the four volumes of test resonators with increasing optical loading is depicted in Fig.$~$\ref{fig:NEPThick}.
The responsivity of the two lowest volume resonators was  greater than desired for 410\,GHz.
At 1\,Hz, the TLS noise is subdominant to the photon noise at the relevant loading levels, though it still has a significant contribution to the overall noise.
For the largest volume resonator we see photon noise limited performance with a power spectral density (PSD) of 2$\times 10^{-17}$, 
while we see TLS noise with a PSD of 1$\times 10^{-17}$ at 1\,Hz.
The expected contribution of the 1/f noise from the inductor at 1\,Hz or lower is observed \cite{Wheeler:22}.
Regardless, the measured detectors over all volumes showed sensitivities less than 10$^{-15}$\,W/Hz$^{0.5}$ (Fig.~7).

\begin{figure*}[!t]
\centering
\includegraphics[width=0.475\linewidth]{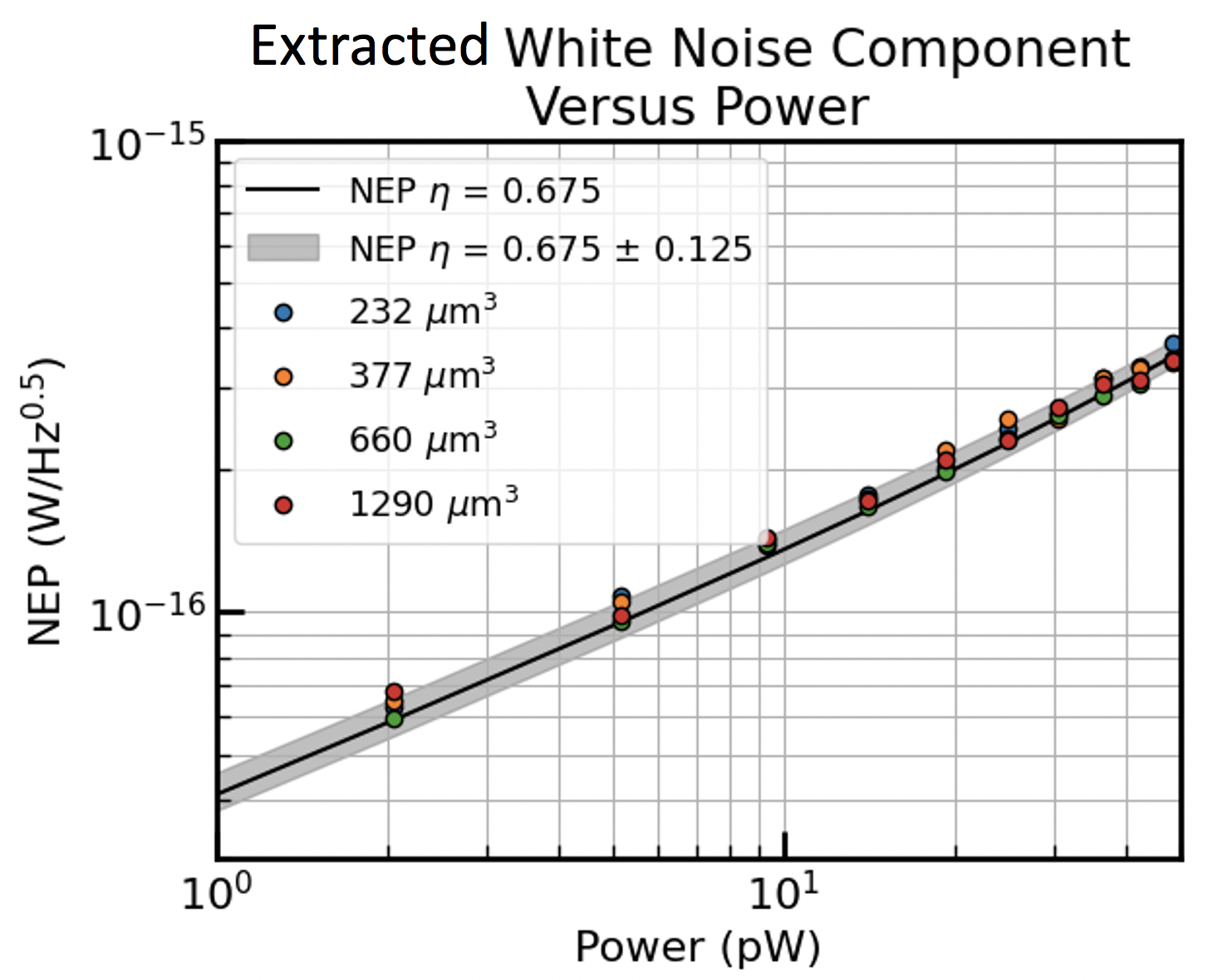}
\includegraphics[width=0.517\linewidth]{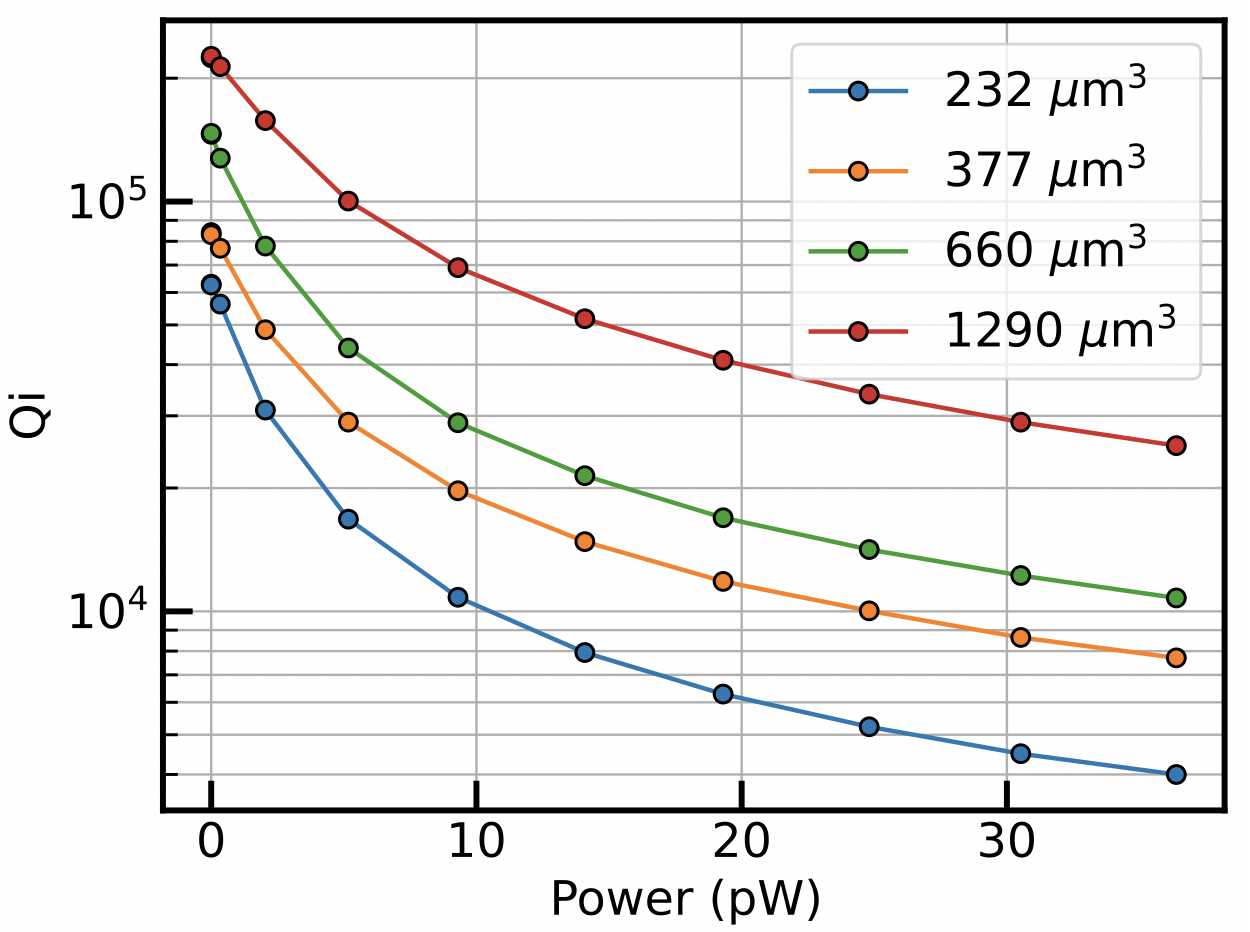}
\caption{Left: White noise equivalent power as a function of incident power for the different volume resonators (252 through 1,290 $\mu$m$^3$) showing the fitted optical efficiency, $\epsilon$. The results of fitting show an optical efficiency of $\sim$0.675 Right: The internal quality factor, $Q_i$, as a function of input power for loading levels expected for the 410\,GHz module. For the expected typical loading level of 10-20\,pW, $Q_i$ is optimal for the 660 $\mu$m$^3$ device.}
\label{fig:WhiteAndQi}
\end{figure*}

The PSD of the noise sources can be characterized with a fit using the following four-parameter model\cite{Pan:23}:
\begin{equation}
	S_{xx} = \frac{A + B f^{-C}}{1 + (2 \pi f D)^2}
	\label{PSD_Fit}
\end{equation}
\noindent where $A$ is the resonator white noise, $Bf^{-C}$ is the $1/\sqrt{f}$ noise (dominated by TLS), and $D$ is the quasiparticle lifetime $\tau$.

The estimated sky load for our horn size and filter band at 410\,GHz is in the range of 10-20\,pW. 
At the lowest-level sky load, the noise should be photon-dominated. 
This is best illustrated (as in Fig.~6) by examining the ratio of
noise from no load 
to 
load across different volumes.
In this sense, the 660 $\mu$m$^3$ provides the ideal volume from these four examples for use at 410\,GHz. 
Further, the responsivity of the two lowest volume test resonators is greater than
desired, while that of the highest volume design is no longer photon
noise limited at the desired $Q_i$ (Fig.~7). 

The next step in the analysis is to extract the optical efficiency of the system, $\epsilon$, from the white noise parameter of the system, or $A$ in Eq.$~$\ref{PSD_Fit}.
To get an accurate representation of the response, the white NEP of the resonator is defined as the squareroot of the ratio of white noise to the optical responsivity, or $NEP_{\rm white}=\sqrt{A/R}$.
These values are then compared to the NEP from photon noise, $NEP_\gamma$, and generation-recombination noise, $NEP_{\rm gr}$.
The results for the four volumes of resonators are shown in the left panel of Figure$~$\ref{fig:WhiteAndQi}, which show an optical efficiency of $\sim$0.675.
Values for the optical efficiency have been consistent across all measured resonator volumes.

Analysis of the noise reveals that the desired sensitivity of at least 10$^{-15}$ W/Hz$^{0.5}$ is achieved at the predicted loading.
Furthermore, the NEP is photon noise dominated {at the readout frequency (500\,Hz)}.
Results from previous attempts to produce highly sensitive detectors show that in order to reduce TLS noise in the thick volume detectors, either the capacitor finger must be separated or the overall capacitor area - and therefore the detector footprint - must increase\cite{Gao,Noroozian:09,Baselmans:22}.
The design of the 410\,GHz detectors has room to further optimize the capacitor to reduce TLS noise.

All four volumes tested show good NEP performance.          
The expected efficiency over the 390 to 430 GHz is expected to be above 90\%. 
A detailed calculation can be found in\cite{austermann2026}. 
As shown in Fig.~6, the 660 $\mu$m$^3$ volume appears to be a good choice with 410\,GHz loading expected ($\sim$15\,pW) with our feedhorn size and filter band at CCAT
The optical responsivity of the  660\,$\mu$m$^3$ volume resonator provides a sufficient responsivity to have minimal noise contributions from the readout, while maintaining high enough resonator quality factor to be able to multiplex high numbers of resonators in the available readout bandwidth without {a lot}  of collisions and cross-talk. 

The trade space between $Q_{\rm i}$ and being strongly photon-noise-limited is illustrated in Fig.~7. 
At the lowest-level atmospheric load, the noise should be photon-dominated (as in Fig.~6).
Figure$~$\ref{fig:WhiteAndQi} shows results for $Q_{\rm i}$ with increasing optical loading, which convey relative stability across the expected range of loading for the module.
The mean $Q_{\rm i}$ within the expected loading range is $\sim$30,000 for the fiducial 660\,$\mu$m$^3$ volume selected for 410\,GHz operation.
The proposed target for performance is to have $Q_{\rm i}  \simeq Q_{\rm c}$ under loading, to avoid amplifier noise beginning to dominate. 
It is proposed that a value of $Q_{\rm c}=$30,000 is adopted for arrays utilizing this fiducial volume resonator design.


\section{Conclusion}

The 410\,GHz module will consist of approximately 20,000 TiN, polarization-sensitive, lumped-element kinetic inductance detectors. The large increase in detector count over the lower frequency Prime-Cam modules will enable high mapping speeds. The millions of dusty galaxies probed by these surveys will have significantly improved dust temperature constraints using the 410\,GHz channel\cite{chapman2003}.
The detectors are being designed to be read out using a multi-octave readout architecture, allowing for approximately double the multiplexing of the lower frequency FYST modules, and comparable to the higher frequency 850\,GHz module \cite{Scott2022}.
The tested septalayer film showed excellent qualities for the sheet inductance and critical temperature, enabling a relatively high volume for the resonators in a small footprint.

Several different volumes of candidate 410\,GHz resonator were tested.
The 410\,GHz MKID prototypes showed high responsivity to thermal and optical loading, and photon limited performance.
These initial test results show that the responsivity of the two lowest volume test resonators was greater than desired, while  the highest volume design is no longer photon noise limited at the desired $Q_i$. 
Overall, the performance of the prototype detectors was highly successful, 
with one of the designs (660 $\mu$m$^3$) very nearly achieving all the specifications at the expected loading.
The 660 $\mu$m$^3$ volume resonator, with relatively minimal modifications, could be adopted in the final design for the 410\,GHz detector array.
The optimization of the final detector design will continue based on the results from these prototypes.

\section*{Acknowledgments}
The CCAT project, FYST and Prime-Cam instrument have been supported by generous contributions from the Fred M. Young, Jr. Charitable Trust, Cornell University, Duke University, and the Canada Foundation for Innovation and the Provinces of Ontario, Alberta, and British Columbia. The construction of the FYST telescope was supported by the Gro{\ss}ger{\"a}te-Programm of the German Science Foundation (Deutsche Forschungsgemeinschaft, DFG) under grant INST 216/733-1 FUGG, as well as funding from Universit{\"a}t zu K{\"o}ln, Universit{\"a}t Bonn, and the Max Planck Institut f{\"u}r Astrophysik, Garching.
The 410 GHz module in Prime-Cam is funded by the Canadian Foundation for Innovation, Innovation Fund 2025 Project 46097 and Canadian provincial matching funds. The completion and deployment of the Prime-Cam instrument with the initial instrument modules is supported by a generous contribution from Alex Gerko, Founder and CEO of XTX Markets.
This 410\,GHz detector development work was also supported by NSERC. We acknowledge research support and laboratory facilities from the NRC-HAA. This research was supported in part by grant NSF PHY-2309135 to the Kavli Institute for Theoretical Physics (KITP) during the fall 2025 residence (SCC).


\bibliography{biblio} 
\bibliographystyle{spiebib} 

\vfill

\end{document}